**HST observations of planetary aurorae, a unique tool to study giant magnetospheres**
L. Lamy[1], R. Prangé[1], S. Badman[2], J. Clarke[3], R. Gladstone[4], W. Pryor[5], J. Saur[6]

Ultraviolet (UV) planetary astronomy is a unique tool to probe planetary environments of the solar system and beyond [1]. But despite a rising interest for new generation giant UV telescopes regularly proposed to international agencies, none has been selected yet, leaving the Hubble Space Telescope (HST) as the most powerful UV observatory in activity. As a result, **the UV initiative set up since HST cycle 21 remains highly valuable**, especially for the **comparative study of planetary aurorae and magnetospheres**, described below.

The Far-UV domain – which concerns STIS, ACS and COS – is indeed a key window to investigate the aurorae of the giant planets and their satellites. These bright and localized emissions are produced by atmospheric neutrals (mainly H and $H_2$ for planets, O for moons) that are collisionally excited by precipitating charged particles, accelerated farther in the magnetosphere [2]. Aurorae are therefore a direct, powerful, diagnosis of the electrodynamic interaction between planetary atmospheres, magnetospheres, moons and the solar wind throughout the heliosphere as well as of the underlying plasma processes at work (acceleration, energy and momentum transfer). Auroral FUV spectro-imaging provide key observables, including the spatial topology and dynamics of active magnetospheric regions, the radiated and precipitated power, and the energy of precipitating electrons. Since 1993, HST regularly observed the auroral emissions of the Jupiter, Saturn and Uranus systems (Fig. 1), leading to significant discoveries and achievements (e.g. planet-satellite interactions, large-scale current systems feeding in auroral ovals, solar wind-induced storms etc.). This rich legacy remains of high interest for further statistical and long-term studies, which can benefit recent open high-level databases such as APIS [3]. But new observations are necessary to comparatively tackle pending questions, under varying solar or seasonal cycles.

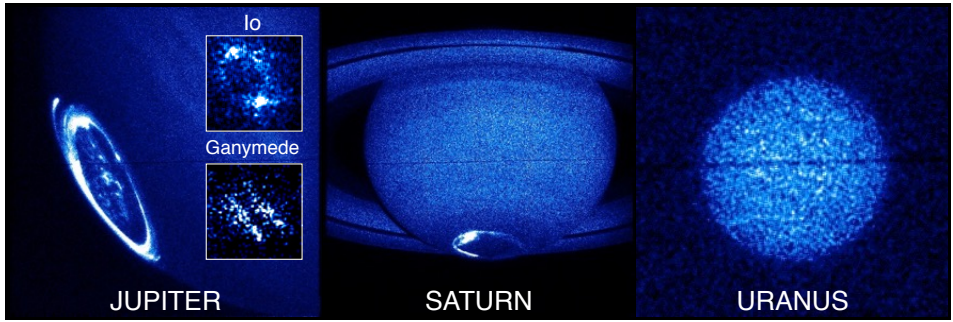

Figure 1 : Collage of HST/STIS FUV auroral observations of the giant planets and their moons.

**Symmetric magnetospheres : Saturn and Jupiter**
In the past, HST often observed Jupiter's and Saturn's circumpolar aurorae in combination with space probes dedicated to planetary exploration such as Galileo (orbiting Jupiter over 1995-2003), Cassini (Jupiter flyby in 2000, orbiting Saturn since 2004) and New Horizons (Jupiter flyby in 2007). Such synergistic observations proved to be essential to assess complex magnetospheric processes with *remote* HST auroral observations simultaneous to *in situ* plasma measurements from spacecraft probing key regions such as the solar wind, the magnetotail etc. [e.g. 4]. **In 2016-2017, HST will have a unique opportunity to observe the aurorae of Saturn and Jupiter while Cassini and Juno[*] orbiters will sample *in situ* their polar regions**, along auroral magnetic field lines up to a few planetary radii above the atmosphere, while also providing *remote* observations of nightside aurorae, invisible to HST.

So-called auroral regions are a key interface between the magnetosphere and the ionosphere, which host fundamental plasma processes suspected to be universal for (exo)planetary magnetospheres, namely (i) particle acceleration and (ii) the generation of radio auroral emission **[5]**. The latter is indeed generated by an electron-wave resonance within plasma cavities where acceleration does take place. The past study of terrestrial auroral regions with numerous polar orbiters funded our current understanding of auroral physics. However, crucial questions remain on the nature of the dominant acceleration process (quasi-static potential drops vs alfvenic acceleration, inter-hemispheric magnetic conjugacy) or on the conditions in which Auroral Kilometric Radiation is generated and propagates out of auroral plasma cavities. Such questions still motivate proposals for auroral multi-point missions **[6]**.

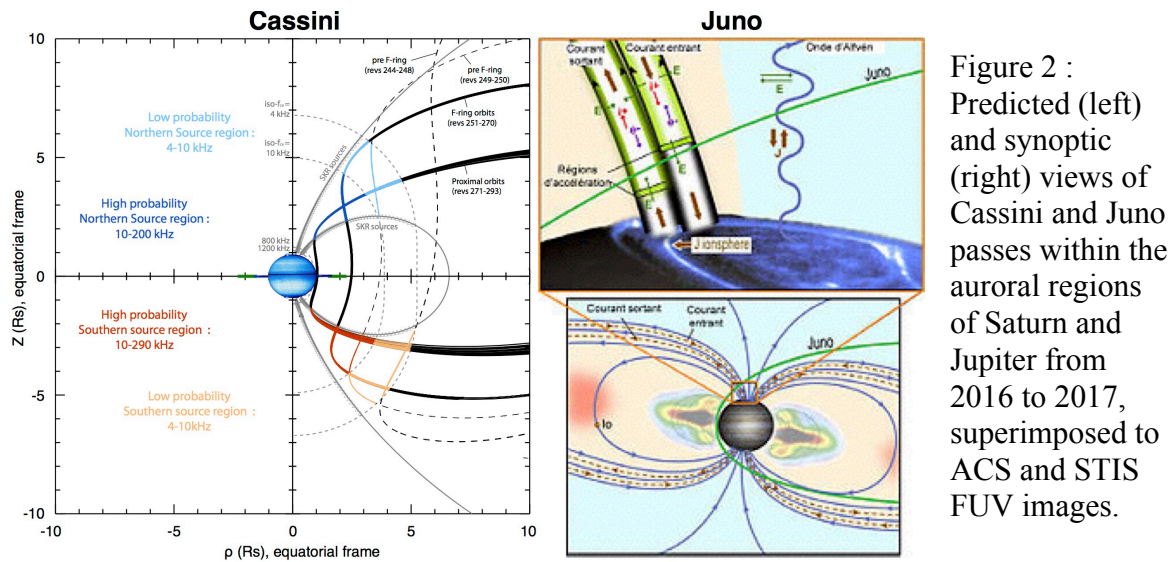

Figure 2 : Predicted (left) and synoptic (right) views of Cassini and Juno passes within the auroral regions of Saturn and Jupiter from 2016 to 2017, superimposed to ACS and STIS FUV images.

Beyond the Earth, only the southern source of Saturn Kilometric Radiation was coincidently crossed once in 2008 by Cassini, in a place where radio sources are generally absent. This first opportunity to investigate the Kronian auroral region revealed preliminary important differences with the terrestrial reference, such as the lack of a plasma cavity. However, this single region was everything but typical, particle measurements were incomplete and no auroral imaging was carried on, preventing any complete analysis **[7]**. In 2016-2017, Cassini will culminate with the latest phase of the mission, *The Grand Finale* : 43 polar orbits flying through previously-unexplored northern and southern auroral regions with a variety of distances, local times and longitudes. Coincidently, the Juno spacecraft, specifically designed to study the auroral regions of Jupiter will at the same time sample them in situ with 32 polar orbits similarly covering a wide range of positions (Fig. 2). Such auroral passes are unlikely to be repeated again in the future. Simultaneous observations by HST are essential to provide the global context for the localized measurements to be acquired by these probes, and combined observations of both planets set an ideal frame for comparative planetology. Only HST can provide the necessary view of the whole auroral region at high spatial resolution (Fig. 1), while its time-tag capability allows the dynamics of all the auroral components to be assessed at timescales down to a few seconds, unachievable to Cassini/Juno spectro-imagers. Nightside measurements of the latter, a few hours after each perikrone/perijove pass, will also be ideally complemented by HST dayside observations to achieve a complete view of aurorae in both hemispheres, necessary to investigate inter-hemispheric magnetic conjugacy.

**Asymmetric magnetospheres : Uranus and Neptune**

Contrary to Saturn and Jupiter, only scarce (almost no) FUV observations of Uranus (Neptune) have been conducted with HST, owing to their distance to Earth and the expected faint level of auroral emissions, while no exploration of ice giant planets is planned yet. Until recently, our knowledge of these sister magnetospheres even remained based on the single observations of Voyager 2, which flew by these planets in 1986 and 1989. At that time, it discovered highly tilted and offset magnetic fields with respect to the spin axis, which produces asymmetric magnetospheres unlike any other one in the solar system. Their configuration with respect to the solar wind flow varies over very different timescales, from a planetary rotation (16-17h) to seasons (decades). Uranus aurorae were identified by the Voyager 2 UV spectrometer as faint nightside H, $H_2$ emissions of a few kR modulated at the planetary rotation. At the prevailing solstice, when the magnetic axis was nearly orthogonal to the solar wind flow, such emissions were proposed to be driven by Earth-like solar wind convection. Neptune's aurorae were even fainter and not unambiguously detected. After two unsuccessful attempts with STIS in 1998 and ACS in 2005, the Uranian aurorae were then redetected with STIS, with the first images of these emissions, through successive campaigns in 2011, 2012 and 2014. To maximize the chances of detection, these campaigns were scheduled during active solar wind conditions rather than randomly [8]. The number of positive detections now reaches a total of 6, only. The transient faint spots detected in 2011, near equinox, were tentatively attributed to dayside reconnection with the interplanetary magnetic field. While subsequent theoretical and modeling works pointed out the difficulty of reconnection to occur [9,10], the analysis (in progress) of the latest detections of longer-lived, extended, auroral spots in 2012 and 2014 seems to confirm this interpretation, with a prominent driving role of interplanetary shocks (i.e. magnetospheric compressions). Overall, these detections opened a new era in the study of asymmetric magnetospheres with numerous emerging questions (plasma source and acceleration, nature and dynamics of the solar wind/magnetosphere mutual interaction, location of magnetic poles, rotation period), and HST is the only tool capable of *remotely* sampling such faint emissions. **It is therefore of primary scientific significance to continue collecting detections of Uranus aurorae with HST on a regular basis while the planet moves from equinox, but also to attempt to similarly detect Neptunian aurorae,** during active solar wind conditions.

**Moons :**

Galilean satellites also host auroral phenomena, induced by the Io-Jupiter type of interaction [11], or by the intrinsic magnetic field of Ganymede (Fig. 1). The limited HST observations of them served to constrain the planet-satellite current systems, Europa water vapor aurorae [12] or the sub-surface salty ocean of Ganymede [13]. Continued HST observations of active moons **remain highly valuable to achieve a more statistical database, complete future HST and Juno observations of Jupiter, and prepare Europa-Clipper and JUICE.**